\documentclass[aip,twocolumn,superscriptaddress,preprintnumbers,showpacs,amsmath,amssymb]{revtex4}
\usepackage{graphicx}
\usepackage[colorlinks=true,citecolor=blue]{hyperref}

\begin{document}
    \title{ Unconventional photon blockade 
    in doubly resonant microcavities \\ with second-order nonlinearity }
    \author{Dario Gerace}
    \email{dario.gerace@unipv.it}
    \affiliation{Department of Physics, University of Pavia, I-27100 Pavia, Italy}
         \author{Vincenzo Savona}
     \affiliation{ Institute of Theoretical Physics, Ecole Polytechnique F\'{e}d\'{e}rale de Lausanne (EPFL), 
                           CH-1015 Lausanne, Switzerland }        

\begin{abstract}
It is shown that non-centrosymmetric materials with bulk second-order nonlinear susceptibility can be used to generate strongly antibunched radiation at an arbitrary wavelength, solely determined by the resonant behavior of suitably engineered coupled microcavities. The proposed scheme exploits the unconventional photon blockade of a coherent driving field at the input of a coupled cavity system, where one of the two cavities is engineered to resonate at both fundamental and second harmonic frequencies, respectively. Remarkably, the unconventional blockade mechanism occurs with reasonably low quality factors at both harmonics, and does not require a sharp doubly-resonant condition for the second cavity, thus proving its feasibility with current semiconductor technology.
\end{abstract}
\maketitle

\textit{Introduction}.
There is currently pressing need for the development of integrated quantum technologies allowing for the generation and manipulation of quantum
states of the electromagnetic radiation, with the ultimate goal of defining a photonic-based architecture for quantum information processing \cite{qp_review}.
For interfacing with long distance infrastructures based on fiber-optics communication, state-of-art sources of quantum radiation have been lately developed at the typical telecommunication wavelengths, either based on heralding photons \cite{Fasel2004,kartik2012} or on artificial quantum emitters \cite{notomi2012scirep}.
However, a source of quantum radiation that is not related to any resonant behavior of a quantum emitter, but can be engineered to operate at arbitrary wavelength and work at room  temperature has not yet been realized. To this end, the single-photon blockade of a strongly nonlinear system can be exploited to convert a coherent radiation source of defined wavelength into antibunched photon streams \cite{imamoglu99}, as recently done in coupled quantum dot-cavity systems \cite{faraon08nphys,volz2012nphot}, with potential implications for the realization of single-photon transistors \cite{chang07np} and interferometers \cite{gerace_josephson}.

It has been recently proposed that single-photon blockade could be achieved in nanostructured cavities either with second- [$\chi^{(2)}$] \cite{arka2013prb} or third-order [$\chi^{(3)}$] \cite{ferretti2012} nonlinear susceptibility, which can be strongly enhanced by diffraction-limited photonic confinement \cite{rivoire2009,matteo_shg_thg}. On the other hand, given the small values of typical nonlinear coefficients of most semiconducting and insulating materials \cite{boyd_book}, an \textit{unconventional photon blockade} (UPB) process could facilitate achieving antibunched light emission from suitably engineered coupled modes \cite{savona10prl}. Such mechanism is based on destructive quantum interference between distinct driven-dissipative pathways \cite{carmichael85,bamba}, and requires a significantly smaller optical nonlinearity than its conventional counterpart. It has been recently proposed that UPB might allow to achieve antibunched light emission either in passive devices made of materials with a large $\chi^{(3)}$ susceptibility, such as silicon \cite{ferretti2013}, or in coupled optomechanical systems \cite{cinesi_optomech,savona2013arxiv}.

In this work we investigate the possibility of achieving UPB in nonlinear materials with $\chi^{(2)}$ susceptibility, following the proposal in Ref. \onlinecite{arka2013prb} to obtain conventional single-photon blockade in photonic microcavities made of a high-$\chi^{(2)}$ nonlinear material - such as, e.g., III-V semiconductors (GaAs, GaP, GaN, AlN, etc.) - and fulfilling a doubly resonant condition - i.e. possessing two confined modes at fundamental and second-harmonic frequencies \cite{johnson2007opex,johnson2012opex,solomon2012preprint}, respectively. 
Since the scheme proposed in Ref. \onlinecite{arka2013prb} posed stringent requirements on the cavity mode quality factor ($Q \sim 10^6 $) and on the doubly resonant condition, here we show that the UPB mechanism allows to significantly relax both those requirements. 
{Moreover, the present $\chi^{(2)}$-based UPB can potentially be achieved with larger values of the effective nonlinear interaction as compared to passive $\chi^{(3)}$ nonlinear devices \cite{ferretti2013}, eventually bringing the overall system parameters closer to the realm of what is reasonably achievable with current technology. }

\begin{figure}[t]
\begin{center}
\includegraphics[width=0.48\textwidth]{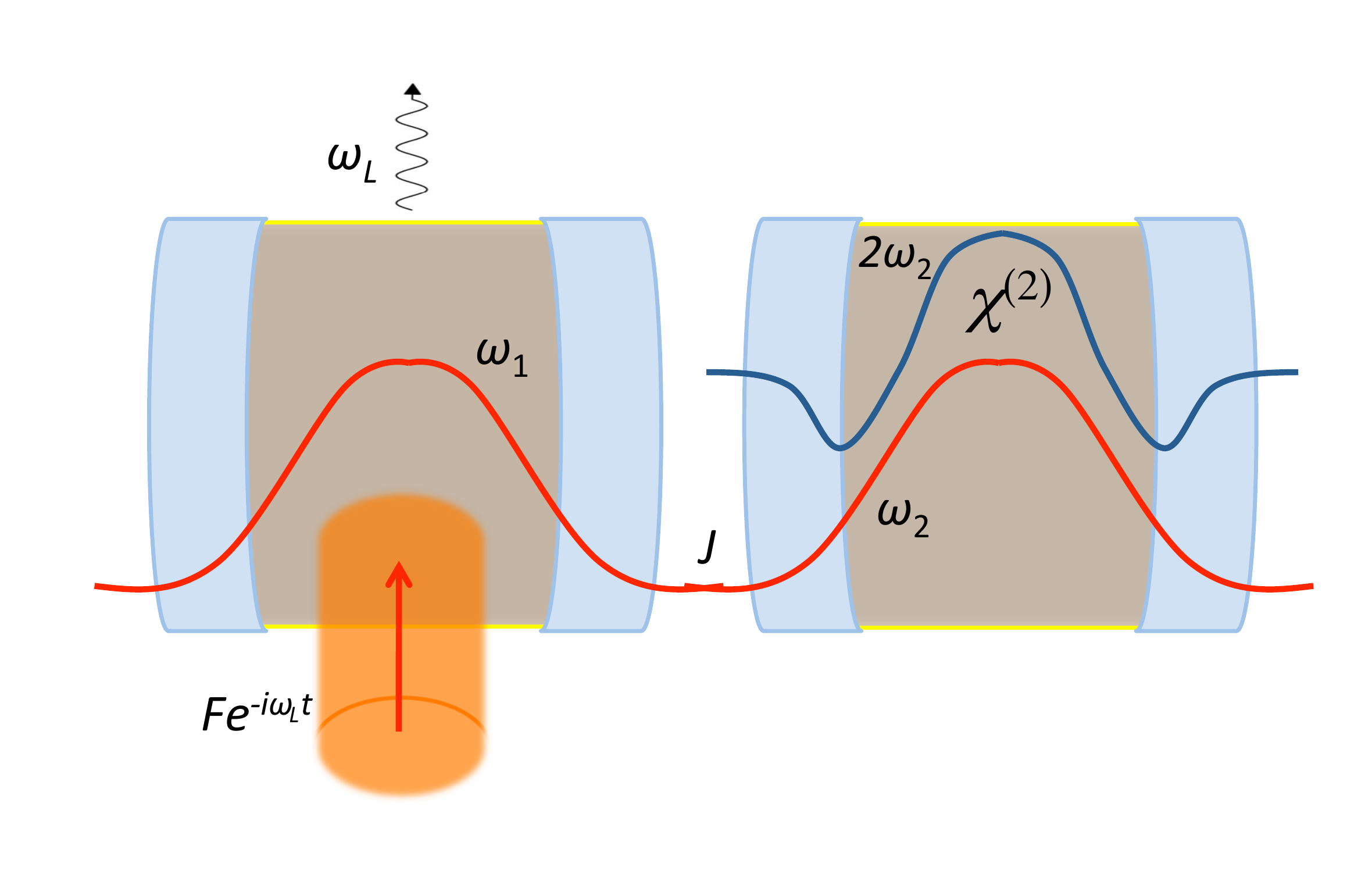}
\vspace{-0.5cm}
\caption{(Color online) Scheme of the system under investigation: two tunnel-coupled resonators ($\omega_1 \sim \omega_2$) 
are driven through the direct injection of a coherent field into cavity 1, while cavity 2 has a doubly 
resonant condition for modes at frequencies $\omega_2$ and $\omega_3 \sim 2 \omega_2$, respectively.
The latter are nonlinearly coupled by a bulk $\chi^{(2)}$ susceptibility. Within this scheme, output of the device is only collected after 
cavity 1. } 
\label{fig1}
\end{center}
\end{figure}

\textit{Theoretical model}. 
We consider a model of driven-dissipative coupled resonators, as schematically described in Fig.~\ref{fig1}.
The system dynamics can be exactly modeled by solving the Liouville-von Neumann master equation for the density matrix 
\begin{equation}\label{eq:master}
\frac{ \mathrm{d} {\rho}}{\mathrm{d}t}={i}[{\rho},\hat{H}] + \mathcal{L}(\rho) \, ,
\end{equation}
where the second-quantized Hamiltonian of the whole system reads \cite{drummond80,carusotto99prb,irvine06prl}
\begin{eqnarray}\label{eq:ham}
\hat{H}&=& \sum_{i=1}^{3} \Delta_{i} \hat{a}_i^{\dagger}\hat{a}_i 
+ J (\hat{a}_{1}^{\dag}\hat{a}_{2}+\hat a_{2}^{\dag}\hat{a}_{1})  \nonumber \\ 
&+& g_{\mathrm{nl}} [\hat{a}_3(\hat{a}_2^\dag)^2+\hat{a}_3^\dag \hat{a}_2^2 ]   
+ F \hat{a}_1^\dag + F^{\ast} \hat{a}_1  \, ,
\end{eqnarray}
assuming $\hbar=1$ throughout this work, for easier notation.
Here, $\hat{a}_1$ ($\hat{a}_2$) describes the quantized fundamental mode in resonator 1 (resonator 2), 
while $\hat{a}_3$ is the second-harmonic mode in the second resonator. The latter is coupled to the $\hat{a}_2$ 
mode by the second-order nonlinear coefficient, $g_{\mathrm{nl}}$,
enhanced by the doubly resonant condition \cite{arka2013prb}, while the nonlinear optical properties of mode 
$\hat{a}_1$ are assumed to be negligible. 
{We notice that previous studies on UPB with Kerr-type nonlinearity have shown that assuming a nonlinear response also 
in the driven cavity does not significantly affect the result \cite{bamba}. As a matter of fact, this consideration holds to a greater 
extent in the present case where, even in presence of a $\chi^{(2)}$-response in both cavities, the driven cavity will not necessarily be 
optimized with a doubly resonant condition. }
Equation (\ref{eq:ham}) is written in a rotated frame with respect to the driving laser frequency, such that
$\Delta_{1,2}=\omega_{1,2}-\omega_L $, and $\Delta_3=\omega_3-2\omega_L$.
The fundamental modes in the two resonators are evanescently coupled to each other 
through the tunneling rate $J$, and the first resonator is pumped at a rate $F$, while losses of the three 
modes are described in Eq. (\ref{eq:master}) by the Lindblad term 
$\mathcal{L} (\rho)= \sum_{i} \kappa_i [\hat{a}_i {\rho} \hat{a}_i^{\dagger} - \hat{a}_i^{\dagger}\hat{a}_i {\rho}/2- {\rho}\hat{a}_i^{\dagger} \hat{a}_i /2]$.
The nonlinear interaction coefficient in the second resonator can be directly determined from the material $\chi^{(2)}$, 
which makes the model suitable to describe resonators made of non-centrosymmetric materials (such as III-V semiconductors). 
In particular, with the simplifying assumption that fundamental and second-harmonic modes have a large spatial overlap \cite{arka2013prb}, 
such term is approximately reduced to 
$g_{\mathrm{nl}} \simeq  \varepsilon_0 \left[{\omega_2 }/({\varepsilon_0 \varepsilon_r })\right]^{3/2} {\bar{\chi}^{(2)}}/{\sqrt{V_{\mathrm{eff}}}}$,
where $\bar{\chi}^{(2)}$ gives a scalar approximation for the second-order susceptibility tensor, $\chi^{(2)}_{ijk}$,
coupling the relevant fundamental and second-harmonic field components, respectively \cite{sipe87prb}.
In the latter expression, $V_{\mathrm{eff}}$ is an effective volume arising from the confinement of the classical field profiles in the resonator, 
assumed to be described by a normalized scalar function $f(\mathbf{r})$ such that $\int |f(\mathbf{r})|^2 \mathrm{d}^3 \mathbf{r} =1 $. Within this
formalism, the effective volume is actually defined as $V_{\mathrm{eff}} = [\int |f(\mathbf{r})|^3 \mathrm{d}^3 \mathbf{r}]^{-2}  $, as previously
derived \cite{arka2013prb}.

In the following, we will study the occurrence of single-photon blockade of an input laser field, described by the last two terms in Eq. (\ref{eq:ham}).
As a relevant figure of merit for photon antibunching \cite{loudon_book}, we will specifically refer to the second-order correlation function at zero time delay from the 
output of the first cavity: 
$g_{11}^{(2)}(0) = \langle\hat{a}_1^{\dag 2}\hat{a}_1^{2}\rangle/ \langle\hat{a}_1^{\dag}\hat{a}_1\rangle^2 =
\mathrm{Tr}\{\hat{a}_1^{\dag2}\hat{a}_1^{2} {\rho}_{ss} \} / n_1^2$,
where $n_1=\mathrm{Tr}\{\hat{a}_1^{\dag}\hat{a}_1 {\rho}_{ss} \}$, and ${\rho}_{ss}$ is the steady state solution corresponding to
$ {\mathrm{d} {\rho}}/{\mathrm{d}t} = 0$ in Eq.~(\ref{eq:master}). 
The main approximations of this model with respect to realistic implementations are as follows. 
First of all, we are assuming that the evanescent coupling of neighboring photonic resonators
can be described within a tight-binding scheme, where the tunnel-coupling rate simply derives from
the overlap between the evanescent tails of the cavity mode profiles (see scheme in Fig.~\ref{fig1}). 
However, care must be taken in the case of photonic crystal cavities at large coupling, where additional 
phases might be added to the tunneling term \cite{caselli2012}.
Then, mixing of input-output channels is neglected here, although it can be straightforwardly taken into account 
within this formalisms \cite{flyac2013}.
Finally, we are neglecting any nonlinear source of losses in Eq.~(\ref{eq:master}), which is justified at low pumping
rates (i.e. low photon occupation in the first cavity, $n_1 \ll 1$). 

\textit{Analytic solution}.
We assume to relax the second-harmonic condition by several linewidths, i.e. $\omega_1=\omega_2$ and $|\omega_3 - 2 \omega_2 | / \kappa \gg 1$, where we will set $\kappa_1=\kappa_2=\kappa$ henceforth.
In the low-pumping limit and for $\omega_L \sim \omega_2$, the nonlinear term in Eq. (\ref{eq:ham}), $\hat{H}_{\mathrm{nl}}= g_{\mathrm{nl}} [\hat{a}_3(\hat{a}_2^\dag)^2+\hat{a}_3^\dag \hat{a}_2^2 ]$, is effectively described by a Kerr-type nonlinear Hamiltonian, i.e.  $\hat{H}_{\mathrm{nl}}^{\prime}= U_{\mathrm{eff}}(\hat{a}_2^\dag \hat{a}_2 + \hat{a}_2^\dag \hat{a}_2^\dag \hat{a}_2 \hat{a}_2)$, with the nonlinear shift $U_{\mathrm{eff}} \simeq g_{\mathrm{nl}}^2 / (\Delta_3-2\Delta_2)$. 
{One simple way of deriving this result is to compute the energy of the state $|0,2,0\rangle$ accounting for the coupling to the state $|0,0,1\rangle$ to lowest-order perturbation theory. Here $\{ | n_1, n_2, n_3  \rangle \}$ are photon number states, where $n_i$ are the occupation numbers of modes 1-3, respectively. }
Given the analogy of this effective model with the one studied in Refs. \onlinecite{savona10prl,bamba}, UPB can be expected to occur in such doubly resonant system even with $g_{\mathrm{nl}} \ll \kappa$, thus relaxing the stringent conditions on the fundamental mode quality factor, $Q_{1,2}=\omega_{1,2} / \kappa$. 
In particular, an analytic solution can be given for the optimal system parameters giving rise to strong antibunching \cite{bamba}: laser frequency detuning, $\Delta_1^{\mathrm{opt}}=\Delta_2^{\mathrm{opt}}=-\kappa/2\sqrt{3}$, and tunnel-coupling rate, $J_{\mathrm{opt}} / \kappa \simeq [(2/3\sqrt{3})\kappa/U_{\mathrm{eff}}]^{1/2}$. In the following, we will set the laser frequency close to the optimal detuning condition, $\Delta_1 = \Delta_2 = - 0.28 \kappa \ll \Delta_3$ (hence $U_{\mathrm{eff}} \simeq g_{\mathrm{nl}}^2 / \Delta_3$), and we will refer all the energy scales to the fundamental mode linewidth (i.e., $\kappa = 1$). We will then study the numerical results by solving the master equation for the full model (\ref{eq:ham}) as a function of the relevant parameters.

\begin{figure}[t]
\begin{center}
\includegraphics[width=0.46\textwidth]{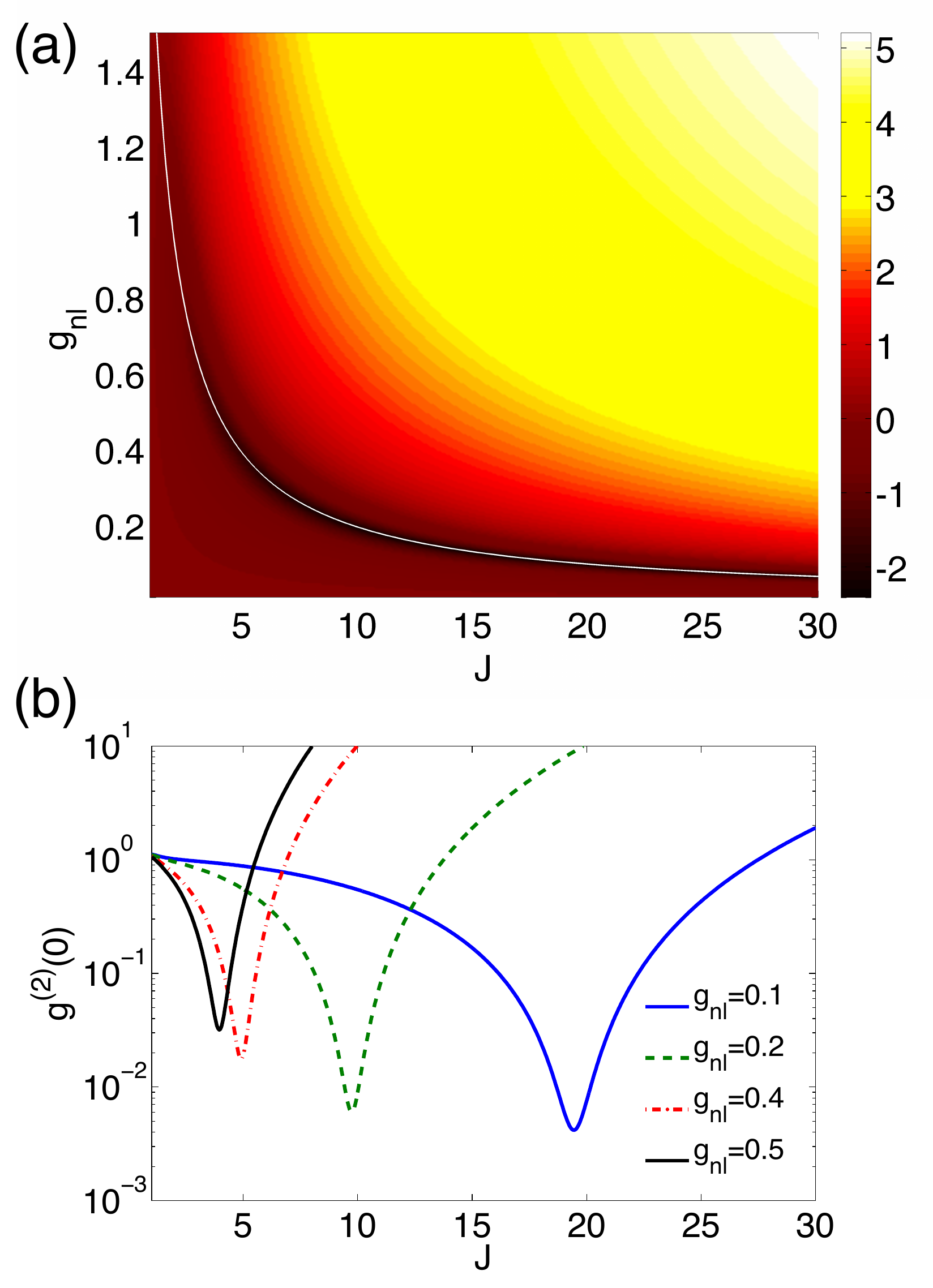}
\caption{(Color online) (a) Color scale plot of $\log_{10} [g_{11}^{(2)}(0)]$ as a function of tunnel coupling ($J$) and second-order nonlinearity ($g_{\mathrm{nl}}$), for parameters (in units of $\kappa$): $F=1$, $\kappa_3 =2 $, $\Delta_1=\Delta_2 = -0.28$, $\Delta_3 = 10$. The white line is a plot of the optimal antibunching condition for the effective Kerr-model (see text). (b) A few cuts taken from the color plot, displaying $g_{11}^{(2)}(0)$ as a function of $J$, for different values of $g_{\mathrm{nl}}$.} \label{fig2}
\end{center}
\end{figure}

\textit{Numerical results}.
From a numerical point of view, the steady state condition of Eq. (\ref{eq:master}) is solved by representing the field operators on the basis of Fock states defined above. The truncation of the Hilbert space is optimized by setting different cut-off occupations in each field ($n_1 \leq 6$, $n_2 \leq 10$, $n_3 \leq 2$ in this case), and carefully checking for numerical convergence against the total number of excitations ($N_{\mathrm{max}}=10$ in this work). In Fig. \ref{fig2}a, the second-order correlation function at zero time delay for radiation emitted from the driven cavity is shown on a log scale color plot, which highlights the antibunching region as a function of $J$ and $g_{\mathrm{nl}}$, respectively. In these calculations, we assumed $\Delta_3 / \kappa = 10$, and similar quality factors for fundamental and second-harmonic modes, $Q_3=\omega_3/ \kappa_3 \simeq Q_1$, i.e. $\kappa_3 = 2 \kappa$. 
The optimal antibunching condition corresponding to the effective Kerr model, namely a plot of the function $J/ \kappa=[3\sqrt{3}g_{\mathrm{nl}}^2 / (2 \kappa \Delta_3) ]^{-1/2}$ in the figure (full line), is faithfully reproduced by the full numerical solution, confirming the occurrence of the UPB mechanism in such a system. As a consequence, strong photon antibunching is obtained also for $g_{\mathrm{nl}} \ll \kappa$, relaxing the requirements of Ref. \onlinecite{arka2013prb}. Simultaneously, given that the nonlinear shift in the effective Kerr model is given by $g_{\mathrm{nl}}^2 / \Delta_3$, the UPB also allows to relax the doubly resonant condition, i.e. strong antibunching will be possible also for $\Delta_3 / \kappa \gg 1$. From the cuts of the color scale plot shown in Fig. \ref{fig2}b, for a value $g_{\mathrm{nl}} / \kappa = 0.1$ the optimal antibunching condition occurs at $J / \kappa \simeq 19.45$, which will be used in the following calculations.


\begin{figure}[t]
\begin{center}
\includegraphics[width=0.42\textwidth]{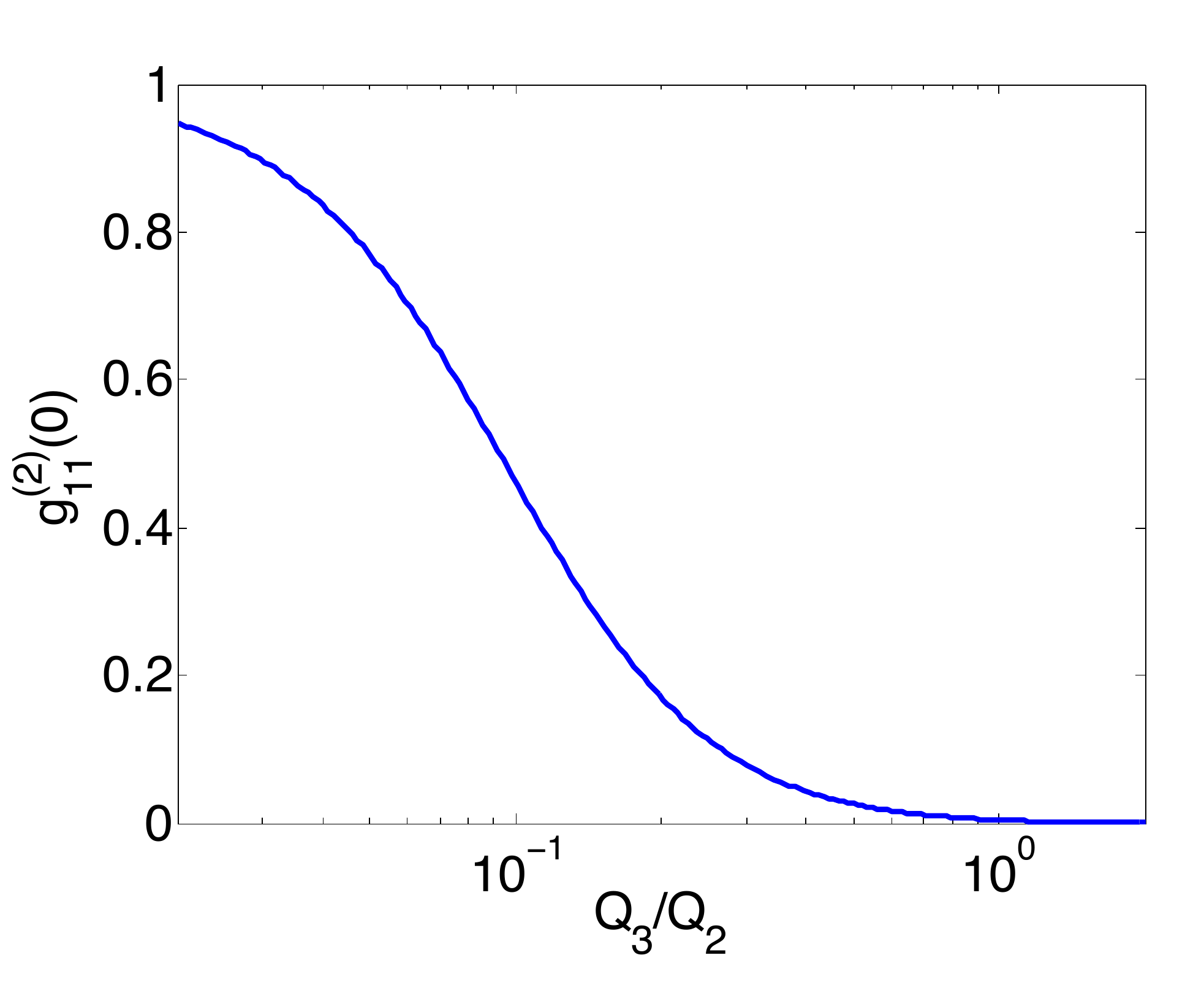}
\caption{(Color online) Dependence of antibunching on the second-harmonic quality factor, $Q_3$, for parameters (in units of $\kappa$): $F=1$, $g_{\mathrm{nl}} = 0.1$, $J = 19.45$, $Q_1=Q_2=Q$. } \label{fig3}
\end{center}
\end{figure}

The next result is shown in Fig. \ref{fig3}, where the dependence of the photon antibunching is checked against the Q-factor of the second-harmonic mode. In particular, this plot is especially relevant in view of the potential difficulties in engineering three-dimensional semiconductor microcavities in which the second-harmonic mode has a comparatively similar Q-factor as the fundamental mode \cite{andreani06prb}. We set the system parameters to the optimal antibunching condition for $g_{\mathrm{nl}} / \kappa = 0.1$, and scan the second-harmonic mode loss rate ranging from $\kappa_3 \simeq \kappa$ (i.e. corresponding to the optimistic condition $Q_3 = 2 Q$ ) up to $\kappa_3\simeq10^2\kappa$. As it is shown in the figure, the antibunching is preserved also for a second-harmonic Q-factor that is an order of magnitude smaller than the fundamental mode. Hence, the second-harmonic Q-factor is less relevant for the UPB mechanism to take place in this system.
UPB is enforced by quantum interference and it is therefore affected by pure dephasing processes occurring in the resonators. More precisely, suppression of photon antibunching occurs when the pure dephasing rate is on the order of the effective nonlinear shift, $U_{\mathrm{eff}}$, as already discussed for the Kerr-type UPB  \cite{ferretti2013}. However, for the passive systems considered in this work, such an effect should be small (mainly determined by thermal fluctuations of the resonances) \cite{rabl2011prl}. 
{ 
Moreover, the pure dephasing of mode 3 can be shown to scarcely affect UPB: a temporally random energy shift $\delta E$ of the eigenvalue corresponding to the state $|0,0,1\rangle$ results (from lowest order perturbation theory, similar to the argument used to derive $U_{\mathrm{eff}}$ above) in a shift $\delta E (g_{nl}/\Delta_3)^2 \ll \delta E$ for the $|0,2,0\rangle$ state. 
}

\begin{figure}[t]
\begin{center}
\includegraphics[width=0.42\textwidth]{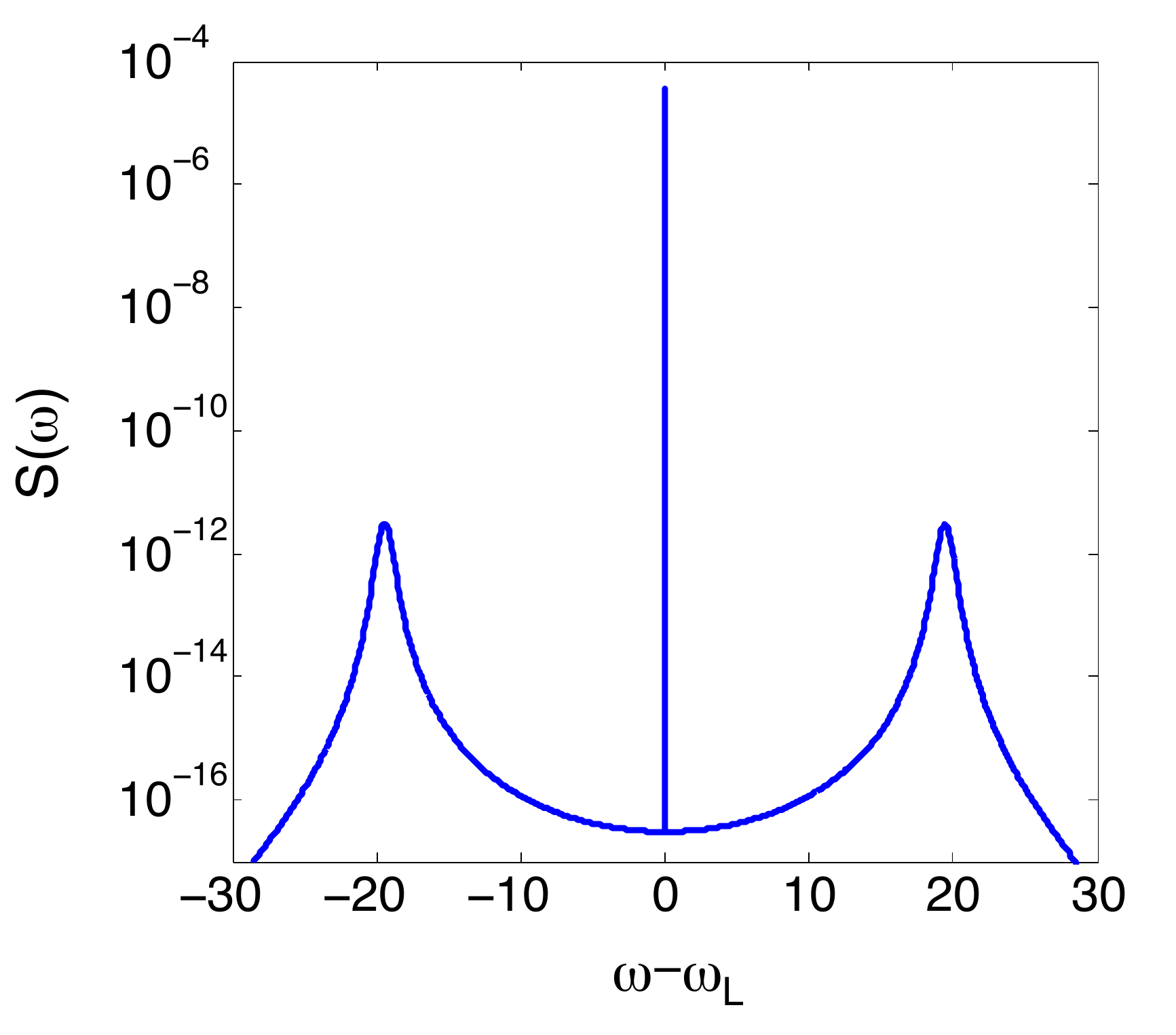}
\caption{(Color online) 
Spectrum of the antibunched radiation emitted from the driven cavity, for parameters (in units of $\kappa$): $F=1$ ($n_1 \simeq 10^{-6}$), $g_{\mathrm{nl}} = 0.1$, $J = 19.45$, and $\kappa_3=2$. } \label{fig4}
\end{center}
\end{figure}



Finally, the spectrum of the emitted photons can be calculated as the Fourier transform $S(\omega)=\int \left\langle \hat{a}^{\dagger}(t) \hat{a}(0)\right\rangle e^{i\omega t } \mathrm{d}t $, and it is shown in Fig. \ref{fig4}b for this UPB source under optimal conditions. The dominant emission evidently occurs at the driving laser frequency. In fact, the peak at $\omega = \omega_{\mathrm{L}}$ inherits the resolution limited linewidth from the (ideally) monochromatic driving field. The external peaks occur at $\omega_1 \pm J$, for $J=19.45$, and they correspond to the normal modes of the coupled cavity system, whose linewidth is instead determined by $\kappa$.  
We notice that although most of the signal will be emitted at frequency $\omega = \omega_{\mathrm{L}}$, it is however a small occupation of the normal modes of the coupled system that produces the destructive quantum interference giving rise to the UPB mechanism \cite{kimble92pra}. 

\textit{Discussion}.
{It is important to stress that the present proposal could be realized with state-of-art technology employing different materials platforms. 
In particular, bulk nonlinear susceptibility can be of the order of $\chi^{(2)}\sim 10-100$ pm/V for the main III-V materials employed in optoelectronics research, such as GaAs \cite{boyd_book,bergfeld2003prl}, GaP, GaN, AlN \cite{chenAPL}. } 
Assuming diffraction-limited cavity confinement, e.g. with engineered photonic crystal cavities,  an estimate of the single-photon nonlinearity for a doubly resonant system with these materials is $g_{\mathrm{nl}} \sim 1$ $\mu$eV \cite{arka2013prb}. Working in the typical telecommunication band, i.e. around $\omega \sim 0.7--0.9$ eV, the required loss rate is $\kappa = 10 g_{\mathrm{nl}} = 10$ $\mu$eV, corresponding to a fundamental mode Q-factor $Q\sim 80000$ (e.g. at 0.8 eV). These values could be routinely achieved in photonic crystal cavities made of III-V semiconductor materials \cite{derossi2008}. 
Interestingly, the required Q-factor for the second-harmonic mode is on the order of 10000, which could stimulate further research in designing doubly resonant microcavities. 
{Finally, for optimal antibunching, the required normal mode splitting in the photonic molecule is $2J \simeq 0.4$ meV with these parameters, but it could be further reduced for larger $g_{\mathrm{nl}}/ \kappa$. As the time interval over which antibunching occurs in UPB is limited by $\pi/J$ \cite{savona10prl,bamba}, this value implies a time resolution of roughly 10 ps for experimentally showing UPB with single-photon correlation measurements \cite{correlations}. }

Among the different types of photonic microcavities, photonic crystal molecules can be fabricated and controlled to a high degree of precision \cite{arka2012prb,caselli2013,waks2013apl}, with a footprint ranging in the few $\mu\mbox{m}$-range, which makes these platforms the one of the preferential system to realize the present proposal in compact and integrated photonic chips. 
{It should be emphasized that in the present scheme both parameters $U_{\mathrm{eff}}$ and $J$ depend on structural details of the coupled cavity system. In particular, $U_{\mathrm{eff}}$ is determined by the detuning $\omega_3-2\omega_2$, as previously discussed. Hence, in a realistic nanofabricated system, the optimal condition $J_{\mathrm{opt}} / \kappa$ will be affected by the tolerance in the fabrication process. For example, the typical uncertainty in the resonant wavelength of a photonic crystal cavity lies within the nm range, while $Q=80000$ corresponds to a cavity linewidth of $\kappa\sim 0.02$ nm. Therefore, device post-selection or post-processing will unavoidably be required for fine tuning, which has already been shown, e.g., in photonic crystal cavities \cite{caselli2013,waks2013apl}.  }

In terms of efficiency, the UPB mechanism is known to be limited to occupancy in the first cavity $n_1 \ll 1$. In fact, the antibunching rapidly degrades for values of $n_1 \geq 10^{-2}$, as already pointed out in Ref. \onlinecite{savona10prl}. By knowing the loss rate $\kappa$ for a specific system implementation, the average photon number allows to estimate the efficiency of this source of antibunched radiation, i.e. the emission rate $R_{\mathrm{em}} = n_1 \kappa$. Assuming a loss rate corresponding to $Q\simeq 8 \times 10^4$ 
at $\omega_{1,2} \simeq 0.8$ eV, one can estimate a maximum antibunched photon rate exceeding $R_{\mathrm{em}} \sim 10$ MHz.

\textit{Conclusion}.
We have shown that antibunched radiation can be obtained at the output of a coupled cavity system under coherent continuous wave driving, only exploiting the bulk material second-order nonlinearity and without the need for quantum emitters or cavity QED effects. The mechanism relies on an unconventional photon blockade induced by quantum interference between excitation/de-excitation pathways, and it is only subject to suitable engineering of the coupled cavity system and the operational conditions. In particular, it is important for at least one of the two resonators to be engineered for a doubly resonant condition at fundamental and second-harmonic frequencies. We have shown that antibunching of the emitted radiation is robust against the second harmonic mode quality factor, as well as  the second-harmonic detuning from the sharp doubly-resonant condition. Moreover, the spectrum of such antibunched radiation is dominated by photons at the driving laser frequency.  
This work can be of interest for the realization of integrated sources of quantum radiation in the telecom band, working in room-temperature quantum photonic circuits, as a promising alternative to the use of single quantum emitters such as semiconductor quantum dots.

\textit{Ackowledgements}.
The authors acknowledge the Swiss National Science Foundation for support through the International Short Visits program, project number IZK0Z2-150900.
D.G. ackowledges partial financial support from the Italian Ministry of University and Research through Fondo Investimenti Ricerca di Base (FIRB) 
``Futuro in Ricerca" project RBFR12RPD1.
We are indebted to  A. Majumdar and M. Minkov for very useful discussions and suggestions.


\end{document}